\documentclass[11pt]{revtex4} %Updated Sep 24, 2009%
\usepackage{amssymb,epsf}
\usepackage{latexsym}
\usepackage{color}
\begin{document}
\title{Interacting HDE and NADE in Brans-Dicke Chameleon Cosmology}
\author{Ahmad Sheykhi} \email{sheykhi@mail.uk.ac.ir}
\affiliation{Department of Physics, Shahid Bahonar University, P.O. Box 76175, Kerman, Iran\\
         Research Institute for Astronomy and Astrophysics of Maragha (RIAAM), Maragha,
         Iran}
\author{Mubasher Jamil}
\email{mjamil@camp.nust.edu.pk} \affiliation{Center for Advanced
Mathematics and Physics, National University of Sciences and
Technology, H-12, Islamabad, Pakistan}

          \begin{abstract}
Motivated by the recent work of one of us \cite{SJ}, we
generalize this work to the case where the pressureless dark
matter and the holographic dark energy do not conserve separately
but interact with each other. We investigate the cosmological
applications of interacting holographic dark energy in Brans-Dicke
theory with chameleon scalar field which is non-minimally coupled
to the matter field. We find out that in this model the phantom
crossing can be constructed if the model parameters are chosen
suitably. We also perform the study for the new agegraphic dark
energy model and calculate some relevant cosmological parameters
and their evolution.
\end{abstract}
\maketitle
\newpage

\section{Introduction\label{Int}}
Among various scenarios to explain the acceleration of the
universe expansion, the holographic dark energy (HDE) and
agegraphic dark energy (ADE) models have got a lot of enthusiasm
recently. These models are originated from some considerations of
the features of the quantum theory of gravity. That is to say, the
HDE and ADE models possess some significant features of quantum
gravity. Although a complete theory of quantum gravity has not
established yet today, we still can make some attempts to
investigate the nature of dark energy according to some principles
of quantum gravity. The former is motivated from the holographic
principle \cite{Li,Huang}. It was shown in \cite{Cohen} that in
quantum field theory, the UV cutoff $\Lambda$ should be related to
the IR cutoff $L$ due to limit set by forming a black hole. If
$\rho_D=\Lambda^4$ is the vacuum energy density caused by UV
cutoff, the total energy of size $L$ should not exceed the mass of
the system-size black hole:
\begin{equation}
E_D\leq E_{BH}\rightarrow L^3\rho_D\leq m_p^2L.
\end{equation}
If the largest cutoff $L$ is taken for saturating this inequality,
we get the energy density of HDE as
\begin{equation}
\rho_D=\frac{3c^2m_p^2}{L^2}=\frac{3c^2}{8\pi G L^2}.
\end{equation}
The HDE is thoroughly investigated in the literature in various
ways (see e.g \cite{Shey0} and references therein ). The later
(ADE) model assumes that the observed dark energy comes from the
spacetime and matter field fluctuations in the universe. Following
the line of quantum fluctuations of spacetime, Karolyhazy et al.
\cite{Kar} discussed that the distance $t$ in Minkowski spacetime
cannot be known to a better accuracy than $\delta{t}=\beta
t_{p}^{2/3}t^{1/3}$ where $\beta$ is a dimensionless constant of
order unity. Based on Karolyhazy relation, Sasakura \cite{Sas}
discussed that the energy density of metric fluctuations of the
Minkowski spacetime is given by  (see also \cite{Maz})
\begin{equation}\label{rho0}
\rho_{D} \sim \frac{1}{t_{p}^2 t^2} \sim \frac{m^2_p}{t^2},
\end{equation}
where $t_{p}$ is the reduced Planck time and $t$ is a proper time
scale. On these basis, Cai \cite{Cai1} proposed the energy density
of the original ADE in the form
\begin{equation}\label{rho11}
\rho_{D}= \frac{3n^2 m_{p}^2}{T^2},
\end{equation}
where $T$ is the age of the universe. Since the original ADE model
suffers from the difficulty to describe the matter-dominated
epoch, the new ADE (NADE) model  was  proposed by Wei and Cai
\cite{Wei2}, while the time scale was chosen to be the conformal
time instead of the age of the universe. The ADE models have
arisen a lot of enthusiasm recently and have examined and studied
in ample detail \cite{age,shey1,shey2,Karami1}.

It is also of great interest to analyze these models in the
framework of Brans-Dicke (BD) gravity. In recent years the BD
theory of gravity got a new impetus as it arises naturally as the
low energy limit of many theories of quantum gravity such as
superstring theory or Kaluza-Klein theory. The motivation for
studying these models in the BD theory comes from the fact that
both HDE and ADE models belong to a dynamical cosmological
constant, therefore we need a dynamical frame to accommodate them
instead of Einstein gravity. The investigation on the HDE and ADE
models in the framework of BD cosmology, have been carried out in
\cite{Pavon2,BD,Sheykhi1,Sheykhi2}. In the present work, we
consider a BD theory in which there is a non-minimal coupling
between the scalar field and the matter field. Thus the action and
the field equations are modified due to the coupling of the scalar
field with the matter. This kind of scalar field usually called
``chameleon" field in the literature \cite{Khoury}. This is due to
the fact that the physical properties of the field, such as its
mass, depend sensitively on the environment. Moreover, in regions
of high density, the chameleon blends with its environment and
becomes essentially invisible to searches for Equivalence
Principle violation and fifth force \cite{Khoury}. Further more,
it was shown \cite{Khoury,Waterhouse} that all existing
constraints from planetary orbits, such as those from lunar laser
ranging, are easily satisfied in the presence of chameleon field.
The reason is that the chameleon-mediated force between two large
objects, such as the Earth and the Sun, is much weaker than one
would naively expect. In particular, it was shown
\cite{Waterhouse} that the deviations from Newtonian gravity due
to the chameleon field of the Earth are suppressed by nine orders
of magnitude by the thin-shell effect. Other studies on the
chameleon gravity have been carried out in \cite{Chameleon}. Our
work differs from that of Ref. \cite{Sheykhi1} in that we assume a
non-minimal coupling between the scalar field and the matter
field. It also differs from that of Ref. \cite{SJ}, in that we
assume the pressureless dark matter and dark energy do not
conserve separately but interact with each other, while the author
of \cite{SJ} assumes that the dark components do not interact with
each other.

\section{HDE in BD theory with Chameleon scalar field\label{HDE}}
We begin with the  BD chameleon theory in which the scalar field is
coupled non-minimally to the matter field via the action
\cite{Das}{
\begin{equation}
 S=\int{
d^{4}x\sqrt{-g}\left(\phi {R}-\frac{\omega}{\phi}g^{\mu
\nu}\partial_{\mu}\phi
\partial_{\nu}\phi-V(\phi) +f(\phi)L_m \right)},\label{act1}
\end{equation}}
where ${R}$ is the Ricci scalar curvature, $\phi$ is the BD scalar
field { with a potential $V(\phi)$. The chameleon
field $\phi$ is} non-minimally coupled to gravity, $\omega$ is the
dimensionless BD parameter. The last term in the action indicates
the interaction between the matter Lagrangian $L_m$ and some
arbitrary function $f(\phi)$ of the BD scalar field. In the limiting
case $f(\phi) =1$, we obtain the standard BD theory.

{ The gravitational field equations derived from the
action (\ref{act1}) with respect to the metric is
\begin{equation}\label{feq}
R_{\mu\nu}-\frac{1}{2}g_{\mu\nu}R=\frac{f(\phi)}{\phi}T_{\mu\nu}+\frac{\omega}{\phi^2}\Big(\phi_\mu
\phi_\nu-\frac{1}{2}g_{\mu\nu}\phi^\alpha\phi_\alpha\Big)+\frac{1}{\phi}[\phi_{\mu;\nu}
-g_{\mu\nu}\Box\phi]-g_{\mu\nu}\frac{V(\phi)}{2\phi}.
\end{equation}
where $T_{\mu\nu}$ represents the stress-energy tensor for the fluid
filling the spacetime which is represented by the perfect fluid
\begin{equation}\label{2aa}
T_{\mu\nu}=(\rho+p)u_\mu u_\nu+pg_{\mu\nu},
\end{equation}
where $\rho$ and $p$ are the energy density and pressure of the
perfect fluid which we assume to be a mixture of matter and dark
energy. Also $u^\mu$ is the four-vector velocity of the fluid
satisfying $u^\mu u_\mu=-1$. The Klein-Gordon equation (or the wave
equation) for the scalar field is
\begin{equation}\label{phi}
\Box\phi=\frac{T}{2\omega+3}\Big(f-\frac{1}{2}\phi
f_{,\phi}\Big)+\frac{1}{2\omega+3}( \phi V_{,\phi}-2V),
\end{equation}
where $T$ is the trace of (\ref{2aa}). } The homogeneous and
isotropic Friedmann-Robertson-Walker (FRW) universe is described by
the metric
\begin{eqnarray}
 ds^2=-dt^2+a^2(t)\left(\frac{dr^2}{1-kr^2}+r^2d\Omega^2\right),\label{metric}
 \end{eqnarray}
where $a(t)$ is the scale factor, and  $k = -1, 0, +1$ corresponds
to open, flat, and closed universes, respectively. Variation of
action (\ref{act1}) with respect to metric (\ref{metric}) for a
universe filled with dust and HDE yields the following field
equations
\begin{eqnarray}
 &&H^2+\frac{k}{a^2}-\frac{\omega}{6}\frac{\dot{\phi} ^2}{\phi^2}+H
\frac{ \dot{\phi}}{\phi}=\frac{f(\phi)}{3\phi}\left(\rho_M+\rho_D\right){+\frac{V(\phi)}{6\phi}},\label{FE1}\\
 &&2\frac{{\ddot{a}}}{a}+H^2+\frac{k}{a^2}+\frac{\omega}{2}\frac{\dot{\phi} ^2}{\phi^2}+2H
\frac{ \dot{\phi}}{\phi}+\frac{\ddot{\phi}}{\phi}
=-\frac{p_D}{\phi}{+\frac{V(\phi)}{2\phi}},\label{FE2}
\end{eqnarray}
where $H=\dot{a}/a$ is the Hubble parameter, $\rho_D$, $p_D$ and
$\rho_M$ are, respectively, the dark energy density, dark energy
pressure and energy density of dust (dark matter). Here, a dot
indicates differentiation with respect to the cosmic time $t$. The
dynamical equation for the scalar field is{
\begin{equation}\label{FE3}
\ddot\phi+3H\dot\phi-\frac{\rho-3p}{2\omega+3}
\Big(f-\frac{1}{2}\phi f_{,\phi}\Big)+\frac{2}{2\omega+3}
\Big(V-\frac{1}{2}\phi V_{,\phi}\Big)=0.
\end{equation}}
We assume the HDE in the chameleon BD theory has the following form
\begin{equation}\label{rho1}
\rho_{D}= \frac{3c^2\phi }{L^2}.
\end{equation}
The motivation idea for taking the energy density of HDE in BD
theory in the form (\ref{rho1}) comes from the fact that in BD
theory we have $\phi\propto G^{-1}$. Here the constant $3c^2$ is
introduced for later convenience and the radius $L$ is defined as
\begin{equation}\label{L}
L=ar(t),
\end{equation}
where the function $r(t)$ can be obtained from the following
relation
\begin{equation}\label{rt}
 \int_{0}^{r(t)}{\frac{dr}{\sqrt{1-kr^2}}}=\int_{0}^{\infty}{\frac{dt}{a}}=\frac{R_h}{a}.
\end{equation}
It is important to note that in the non-flat universe the
characteristic length which plays the role of the IR-cutoff is the
radius $L$ of the event horizon measured on the sphere of the
horizon and not the radial size $R_h$ of the horizon. Solving Eq.
(\ref{rt}) for the general case of the non-flat FRW universe, we
get
\begin{equation}
r(t)=\frac{1}{\sqrt{k}}\sin y,\label{rt}
\end{equation}
where $y=\sqrt{k} R_h/a$. Now we define the critical energy
density, $\rho_{\mathrm{cr}}$, and the energy density of the
curvature, $\rho_k$, as
\begin{eqnarray}\label{rhocr}
\rho_{\mathrm{cr}}=3\phi H^2,\hspace{0.8cm}
\rho_k=\frac{3k\phi}{a^2}.
\end{eqnarray}
As usual, the fractional energy densities are defined as
\begin{eqnarray}
\Omega_M&=&\frac{\rho_M}{\rho_{\mathrm{cr}}}=\frac{\rho_M}{3\phi
H^2}, \label{Omegam} \\
\Omega_k&=&\frac{\rho_k}{\rho_{\mathrm{cr}}}=\frac{k}{H^2 a^2},\label{Omegak} \\
\Omega_D&=&\frac{\rho_D}{\rho_{\mathrm{cr}}}=\frac{c^2}{H^2L^2}.
\label{OmegaD}
\end{eqnarray}
For latter convenience we rewrite Eq. (\ref{OmegaD}) in the form
\begin{eqnarray}
HL=\frac{c}{\sqrt{\Omega_D}}. \label{HL}
\end{eqnarray}
Taking derivative with respect to the cosmic time $t$ from Eq.
(\ref{L}) and using Eqs. (\ref{rt}) and (\ref{HL}) we obtain
\begin{eqnarray}
\dot{L}=HL+a\dot{r}(t)=\frac{c}{\sqrt{\Omega_D}}-\cos y.
\label{Ldot}
\end{eqnarray}
Consider the FRW universe filled with dark energy and pressureless
matter which evolves according to their conservation laws
\begin{eqnarray}
&&\dot{\rho}_D+3H\rho_D(1+w_D)=0,\label{consq}\\
&&\dot{\rho}_M+3H\rho_M=0, \label{consm}
\end{eqnarray}
where $w_D$ is the equation of state parameter of dark energy. At
this point our system of equations is not closed and we still have
freedom to choose one. We shall assume that BD field can be
described as a power law of the scale factor, $\phi\propto
a^\alpha$. In principle there is no compelling reason for this
choice. However, it has been shown that for small $\alpha$ it
leads to consistent results \cite{Pavon2}. A case of particular
interest is that when $\alpha$ is small whereas $\omega$ is high
so that the product $\alpha \omega$ results of order unity
\cite{Pavon2}. This is interesting because local astronomical
experiments set a very high lower bound on $\omega$ \cite{Will};
in particular, the Cassini experiment implies that $\omega>10^4$
\cite{Bert,Aca}. Taking the derivative with respect to time of
relation $\phi\propto a^\alpha$ we get
\begin{eqnarray}\label{dotphi}
&&\dot{\phi}=\alpha H\phi, \\
&&\ddot{\phi}=\alpha^2H^2\phi+\alpha\phi\dot{H}.\label{ddotphi}
\end{eqnarray}
Taking the derivative  of Eq. (\ref{rho1}) with respect to time
and using Eqs. (\ref{Ldot}) and (\ref{dotphi}) we reach
\begin{eqnarray}
\dot{\rho}_D=H\rho_D\left(\alpha-2+2\frac{\sqrt{\Omega_D}}{c}\cos
y\right)\label{rhodot}.
\end{eqnarray}
Substituting this equation in Eq. (\ref{consq}), we obtain the
equation of state parameter
\begin{eqnarray}
w_D=-\frac{1}{3}(\alpha+1)-\frac{2\sqrt{\Omega_D}}{3c}\cos
y\label{wD}.
\end{eqnarray}
It is important to note that in the limiting case  $\alpha=0$
($\omega\rightarrow\infty$), the Brans-Dicke scalar field becomes
trivial and Eq. (\ref{wD}) reduces to its respective expression in
Einstein gravity \cite{Huang}
\begin{eqnarray}
w_D=-\frac{1}{3}-\frac{2\sqrt{\Omega_D}}{3c}\cos y\label{wDstand}.
\end{eqnarray}
We will see that the combination of the Brans-Dicke field and HDE
brings rich physics. For $\alpha\geq 0$, $w_D$ is bounded from
below by
\begin{eqnarray}
w_D=-\frac{1}{3}(\alpha+1)-\frac{2\sqrt{\Omega_D}}{3c}\label{wDbound}.
\end{eqnarray}
Assuming $\Omega_D= 0.73$ for the present time and choosing $c=1$
\cite{c}, the lower bound becomes $ w_D=-\frac{\alpha}{3}-0.9$.
Thus for $\alpha\geq0.3$ we have $w_D\leq-1$. This implies that
the phantom crossing can be constructed in the BD framework. We
can also obtain the deceleration parameter
\begin{eqnarray}
q=-\frac{\ddot{a}}{aH^2}=-1-\frac{\dot{H}}{H^2},
\end{eqnarray}
which combined with the Hubble parameter and the dimensionless
density parameters form a set of useful parameters for the
description of the astrophysical observations. Dividing  Eq.
(\ref{FE2}) by $H^2$, and using Eqs. (\ref{rho1}), (\ref{HL}),
(\ref{dotphi}) and (\ref{ddotphi}), we find
\begin{eqnarray}\label{q}
q=\frac{1}{\alpha+2}\left[(\alpha+1)^2+
\alpha\left(\frac{\alpha\omega}{2}-1\right)+\Omega_k+3\Omega_D
w_D{-\frac{3}{2}\Omega_{V}}\right]\label{q1}.
\end{eqnarray}
{where the last term can be understood as a
contribution of the potential energy in the total energy density
i.e.
\begin{equation}\label{Ov}
\Omega_V=\frac{V}{\rho_{\mathrm{cr}}}.
\end{equation}}
Substituting $w_D$ from Eq. (\ref{wD}) in (\ref{q}), we get
\begin{eqnarray}
q=\frac{1}{\alpha+2}\left[(\alpha+1)^2+\alpha\left(\frac{\alpha\omega}{2}-1\right)+\Omega_k-(\alpha+1)\Omega_D-\frac{2}{c}{\Omega^{3/2}_D}\cos
y{-\frac{3}{2}\Omega_{V}} \right]\label{q2}.
\end{eqnarray}
If we take $\Omega_D= 0.73$ and $\Omega_k\approx 0.01$  for the
present time and choosing $c=1$, $\alpha\omega\approx1$,
$\omega=10^4$ and $\cos y\simeq 1$, we obtain $q=-0.48$ for the
present value of the deceleration parameter which is in good
agreement with recent observational results \cite{Daly}.

{From equation (\ref{FE3}), we can also estimate the
mass of the chameleon field. This can be done by calculating the
second derivative of the potential function with respect to scalar
field \cite{sergei}. We get
\begin{equation}\label{vf1}
m_\phi^2\equiv V_{,\phi\phi}=\frac{1}{\phi}\Big[
V_{,\phi}-\frac{\rho-3p}{2}(f_{,\phi}-\phi f_{,\phi\phi}) \Big].
\end{equation}
Following previous studies \cite{Das,sergei}, we choose
\begin{equation}\label{vf}
V(\phi)=\frac{M^{4+n}}{\phi^n},\ \ \ f(\phi)=f_0e^{b_0\phi}.
\end{equation}
Here $M$, $f_0$ and $b_0$ are finite parameters whose values are
model dependent. Making use of Eq. (\ref{vf}) in (\ref{vf1}), we
obtain
\begin{equation}\label{vf2}
m_\phi^2=-\frac{1}{\phi}\Big[n\frac{M^{4+n}}{\phi^{n+1}}+\frac{b_0f_0e^{b_0\phi}}{2}(\rho-3p)
(1-b_0\phi) \Big].
\end{equation}
Clearly when $n\rightarrow0$ (which corresponds to a constant
potential), the mass of the scalar field will be dependent on the
properties of $f(\phi)$. Moreover if only $\phi=1/b_0$, the mass is
determined by the scalar potential function alone.}
%%%%%%%%%%%%%%%%%%%%%%%%%%%%%%%%%%%%%%%%%%%%%%%%%%%%%%%%%
\section{Interacting HDE in BD theory with Chameleon scalar field\label{INTHDE}}
In this section we would like to construct a cosmological model
based on the BD chameleon field theory of gravity and on the
assumption that the dark energy and dark matter do not conserve
separately but interact with each other. Taking the interaction
into account the continuity equations  becomes
\begin{eqnarray}
&&
\dot{\rho}_D+3H\rho_D(1+w_D)=-Q,\label{consq2}\\
&&\dot{\rho}_{M}+3H\rho_{M}=Q, \label{consm2}
\end{eqnarray}
where $Q$ is an interaction term which can be an arbitrary
function of cosmological parameters like the Hubble parameter and
energy densities. The dynamics of interacting dark energy models
with different interaction terms have been investigated in
\cite{Cabral}. It should be noted that the ideal interaction term
must be motivated from the theory of quantum gravity. In the
absence of such a theory, we rely on pure dimensional basis for
choosing an interaction $Q$. Hence following \cite{Kim06}, we
assume $Q=\Gamma\rho_{D}$ with $\Gamma=3b^2(1+r)H$ where
$r={\rho_M}/{\rho_{D}}$ is the  ratio of energy densities and
$b^2$ is a coupling constant. Note that $\Gamma>0$ shows that
there is an energy transfer from the dark energy to dark matter.
Combining Eqs. (\ref{rhocr}) and (\ref{dotphi}) with the first
Friedmann equation (\ref{FE1}), we can rewrite this equation as
\begin{eqnarray}\label{Fried2new}
1+\Omega_k=f(\phi)(\Omega_{M}+\Omega_D)+\Omega_{\phi}{+\frac{1}{2}\Omega_{V}},
\end{eqnarray}
where
\begin{eqnarray}\label{Omegaphi}
\Omega_{\phi}=\alpha \left(\frac{\alpha \omega}{6}-1\right).
\end{eqnarray}
Combining Eqs. (\ref{rhodot}),  (\ref{Fried2new}) and
(\ref{Omegaphi}) with  Eq. (\ref{consq2}) we find the equation of
state parameter of the interacting HDE
\begin{eqnarray}
w_D=-\frac{1}{3}(\alpha+1)-\frac{2\sqrt{\Omega_D}}{3c}\cos
y-\frac{b^2}{f(\phi)\Omega_D}\left[1+\Omega_k+\alpha\left(1-\frac{\alpha
\omega}{6}\right){-\frac{1}{2}\Omega_{V}}\right]\label{wDInt}.
\end{eqnarray}
In the absence of the BD  field ($\alpha=0$,  $f(\phi)=1$,
{$V(\phi)=0$}), Eq. (\ref{wDInt}) restores its
respective expression in non-flat standard cosmology \cite{wang2}
\begin{eqnarray}
w_D=-\frac{1}{3}-\frac{2\sqrt{\Omega_D}}{3c} \cos y-\frac{b^2}{
\Omega_D}\left(1+\Omega_k\right)\label{wDIntstand}.
\end{eqnarray}
Next, we examine the deceleration parameter, $q=-\ddot{a}/(aH^2)$.
Substituting $w_D$ from Eq. (\ref{wDInt}) in Eq. (\ref{q1}), one
can easily show
\begin{eqnarray}
q&=&\frac{1}{\alpha+2}\left[(\alpha+1)^2+\alpha\left(\frac{\alpha\omega}{2}-1\right)
+\Omega_k-(\alpha+1)\Omega_D-\frac{2}{c}{\Omega^{3/2}_D}\cos
y{-\frac{3}{2}\Omega_{V}} \right. \nonumber\
\\
&& \left.-\frac{3b^2}{f(\phi)}
\left(1+\Omega_k+\alpha\left(1-\frac{\alpha\omega}{6}\right){-\frac{1}{2}\Omega_{V}}\right)\right]\label{q2hInt}.
\end{eqnarray}
Comparing Eq. (\ref{q2hInt}) with (\ref{q2}) shows that in the
presence of interaction the chameleon function $f(\phi)$ enters
explicitly in $q$ expression. This is in contrast to the
 usual BD theory where $q$ of the interacting HDE model does not depend
on the scalar field \cite{Sheykhi1}.

Finally we present the equation of motion of the dark energy.
Taking the derivative of Eq. (\ref{OmegaD}) and using Eq.
(\ref{Ldot}) and relation ${\dot{\Omega}_D}=H{\Omega'_D}$, we find
\begin{eqnarray}\label{OmegaD2}
{\Omega'_D}=2\Omega_D\left(-\frac{\dot{H}}{H^2}-1+\frac{\sqrt{\Omega_D}}{c}\cos
y \right),
\end{eqnarray}
where the dot is the derivative with respect to time and the prime
denotes the derivative with respect to $x=\ln{a}$. Using relation
$q=-1-\frac{\dot{H}}{H^2}$, we have
\begin{eqnarray}\label{OmegaD3}
{\Omega'_D}=2\Omega_D\left(q+\frac{\sqrt{\Omega_D}}{c}\cos y
\right),
\end{eqnarray}
where $q$ is given by Eq. (\ref{q2hInt}). This equation describes
the evolution behavior of the interacting HDE in BD cosmology with
chameleon field.
\section{Interacting NADE with Chameleon scalar field\label{INTHDE}}
The above study can also be performed for the new agegraphic dark
energy (NADE) model. In NADE, the infrared cut-off is the
conformal time which is defined as
\begin{equation}
\eta=\int\frac{{\rm d}t}{a}=\int_0^a\frac{{\rm
d}a}{Ha^2}.\label{eta}
\end{equation}
In the framework of BD chameleon scalar field, we assume the
following form for the energy density of the NADE
\begin{equation}\label{rho1n}
\rho_{D}= \frac{3n^2\phi }{\eta^2}.
\end{equation}
where the numerical factor $3n^2$ is introduced to parameterize
some uncertainties, such as the species of quantum fields in the
universe, the effect of curved space-time and so on. The
respective fractional energy densities can be written as
\begin{eqnarray}
\Omega_D=\frac{\rho_D}{\rho_{\mathrm{cr}}}=\frac{n^2}{H^2\eta^2}
\label{OmegaDn}.
\end{eqnarray}
Differentiating Eq. (\ref{rho1n}) and using Eqs. (\ref{dotphi})
and (\ref{OmegaDn}) we have
\begin{eqnarray}
\dot{\rho}_D=H\rho_D\left(\alpha-\frac{2}{na}\sqrt{\Omega_D}\right)\label{rhodotn}.
\end{eqnarray}
Substituting this relation in Eq. (\ref{consq2}) and using
relations (\ref{Fried2new}) and (\ref{Omegaphi}), we obtain the
equation of state parameter of the interacting NADE
\begin{eqnarray}
w_D&=&-1-\frac{1}{3}\alpha+\frac{2}{3na}\sqrt{\Omega_D}
-\frac{b^2}{f(\phi)\Omega_D}\left[1+\Omega_k+\alpha
\left(1-\frac{\alpha\omega}{6}\right){-\frac{1}{2}\Omega_{V}}\right]\label{wDnInt}.
\end{eqnarray}
When  $\alpha=0$, $f=1$ { and $V=0$}, the BD scalar
field becomes trivial and Eq. (\ref{wDnInt}) reduces to its
respective expression in NADE in Einstein gravity \cite{shey1}. From
Eq. (\ref{wDnInt}), we see that even in the absence of interaction
$(b=0)$, the the phantom crossing will take place in the the
framework of BD theory provided the model parameters are chosen
suitably. Indeed in this case $(b=0)$,  $w_D$ can cross the phantom
divide provided $na\alpha>2\sqrt{\Omega_D}$. If we take
$\Omega_D=0.73$ and $a=1$ for the present time, the phantom-like
equation of state can be accounted if $n\alpha>1.7$. For instance,
for $n=4$ and $\alpha=0.5$, we get $w_D=-1.02$. When the interaction
is taken into account the phantom crossing for $w_D$ can be more
easily achieved for than when resort to the Einstein field equations
is made.

In the context of BD chameleon scalar field the deceleration
parameter of interacting NADE is obtained as
\begin{eqnarray}
q&=&\frac{1}{\alpha+2}\left[(\alpha+1)^2+\alpha\left(\frac{\alpha\omega}{2}-1\right)
+\Omega_k-(\alpha+3)\Omega_D+\frac{2}{na}{\Omega^{3/2}_D}{-\frac{3}{2}\Omega_{V}}\right.
\nonumber\
\\
&& \left.-\frac{3b^2}{f(\phi)}
\left(1+\Omega_k+\alpha\left(1-\frac{\alpha\omega}{6}\right){-\frac{1}{2}\Omega_{V}}
\right)\right]\label{q2Int}.
\end{eqnarray}
While the equation of motion for $\Omega_D$ takes the form
\begin{eqnarray}\label{OmegaD3n}
{\Omega'_D}=2\Omega_D\left(1+q-\frac{\sqrt{\Omega_D}}{na}\right).
\end{eqnarray}
%%%%%%%%%%%%%%%%%%%%%%%%%%%%%%%%%%%%%%%%%%%%%%%%%%%%%%%%%

%%%%%%%%%%%%%%%%%%%%%%%%%%%%%%%%%%%%%%%%%%%%%%%%%%%%%%%%%%%%%%%%%%
\section{Conclusions\label{CONC}}
In this paper, we have considered interacting HDE model in the
framework of BD cosmology where the HDE density $\rho_{D}=
{3c^2}/(8\pi G L^{2})$ is replaced with $\rho_{D}= {3c^2\phi }/{
L^2}$. With this replacement in BD theory, we found that the
cosmic acceleration will be more easily achieved for than when the
standard HDE is considered. Following the work of \cite{SJ}, we
assumed that the scalar field is non-minimally coupled with the
matter field via an arbitrary coupling function $f(\phi)$. In
principle, the coupling between  BD scalar field and matter field
should be derived from a theory of quantum gravity. In the absence
of such a theory, we have kept our analysis general regardless of
the specification of $f(\phi)$. In the present paper, we have
extended the work \cite{SJ} by incorporating the interaction term
in the HDE model. An interesting consequence of the present model
is that it allows the phantom crossing of the equation of state of
dark energy due to the presence of several free parameters.  We
have also performed the analysis for the NADE model and calculate
some relevant cosmological parameters such as the equation of
state, deceleration parameter and energy density parameter.
%%%%%%%%%%%%%%%%%%%%%%%%%%%%%%%%%%%%%%%%%%%%%%%%%%%%%%%%%%%%%%%%%%%%%%%
\acknowledgments{We thank the anonymous referee for constructive
comments. This work has been supported by Research Institute for
Astronomy and Astrophysics of Maragha, Iran. {M. Jamil would like to
thank E. Abdalla, K. Karami, D. Pavon and S. Odintsov for useful
communications during this work.}}


\begin{thebibliography}{99}
\bibitem{c} Since $\omega\geq10^4$ we find that $\alpha\approx
1/\omega=10^{-4}$, thus Eq. (27) reduces practically to $
w_D=-\frac{1}{3}(1+\alpha)-\frac{2\sqrt{\Omega_D}}{3c}\approx
-\frac{1}{3}-\frac{2\sqrt{\Omega_D}}{3c}$ which is exactly the
Li's result. Thus, the Li's argument \cite{Li} in favor of $c=1$
holds here.
\end{thebibliography}
\end{document}